# Nucleation of Al Nanocrystals in Solute-Substituted Al Metallic Glasses I: Structural characterization

Feng Yi, S. D. Imhoff, J. H. Perepezko and P. M. Voyles

Department of Materials Science and Engineering, University of Wisconsin-Madison, WI, USA

**Abstract**

Primary crystallization in high Al-content metallic glasses is driven by nanometer-diameter regions with internal structure similar to fcc Al. Comparison of fluctuation electron microscopy (FEM) data to FEM simulations of fcc Al clusters dispersed in a dense-random packed matrix is used to extract the diameter and volume fraction of the ordered regions in a $Al_{88}Y_7Fe_5$ base glass and in glasses with 1 at.% Cu substituted for Y or Al. The size and density of nanocrystals were measured as a function of isothermal annealing time for the same alloys. The volume fraction of crystalline material grows under isothermal annealing, so the phase transformation is not purely grain coarsening, but the crystalline volume fraction is lower than the volume fraction of ordered regions in the as-quenched samples, so not all of the ordered regions act as nuclei. Changes in diameter and volume fraction of the ordered regions with alloying are correlated with changes in the crystallization temperature, nucleation rate, and nanocrystal density. No evidence for phase separation is observed, and FEM simulations from a molecular dynamics quenched structural model of similar composition do not show the features observed in experiment.

## 1. Introduction

The primary crystallization reaction in high Al content metallic glasses [1, 2] is interesting for fundamental studies of nucleation. Nucleation occurs at high density [3], but only limited growth of the Al phase occurs due to impingement of the diffusion field of solute expelled by the growing pure Al nanocrystal [4]. This enables experimental measurements of the nucleation density and nucleation rate as a function of time and temperature that are difficult in other systems. The amorphous/nanocrystal composite formed by primary crystallization also has attractive mechanical properties [5-7], and it has similar microstructure to other systems of interest for applications, such as ductile bulk metallic glass (BMG) composites [8] and soft-magnetic Fe-based composites [9]. It is therefore of significant interest to control this structure. Inoue [7] summarized the general pathway to produce Al alloys with nanocrystal dispersions and pointed out the Al amorphous-nanocrystal composite exhibits superior mechanical properties compared to single phase alloy. Foley *et al*. [3] discussed strategies to control the development of nanocrystallites by alloying plus suitable processing routes.

Primary crystallization in Al-based metallic glasses is characterized by an incubation time [10-12] and high composition sensitivity [13, 14]. Transient heterogeneous nucleation was suggested as the mechanism driving the transformation [12], but the heterogeneous nucleation site was not identified. Using isothermal DSC, Nakazato *et al*.[13] found that crystallization of $Al_{85}Ni_{10}Ce_5$ involved nucleation and growth, while crystallization of $Al_{87}Ni_{10}Ce_3$ involved growth only, demonstrating that primary crystallization is very sensitive to the alloy composition.

Three theories have been proposed to explain this phase transformation. Perepezko and Hebert [15] categorized Al-based glass as a growth controlled glass. In this type of glass, during the quenching, the cooling curve on a time-temperature-transformation (TTT) diagram bypasses the growth curve, but intercepts the nucleation curve. Therefore, some seeds nucleate, but due to the dramatic increase of viscosity, their growth is halted by quenching. These clusters are called quenched-in nuclei, and may act as seeds for primary nanocrystalization. Xing *et al*.[16] suggested extremely fine α-Al nanocrystal embedded in amorphous matrix in the as-quenched sample. This is a similar microstructure to quenched-in nuclei, but with the key difference that the phase transformation occurs by constant volume fraction coarsening, not by growth at the expense of aluminum from the matrix.. Unlike a nucleation + growth reaction, there is no delay time in grain coarsening, and the total volume fraction of the Al phase is conserved. The third theory is based on phase separation. Gangopadhyay *et al*. [17] reported that phase separation occurs in the amorphous phase, and that nucleation occurs at the phase interface, which is evidenced by contrast in a TEM of annealed sample. Sahu *et al*. [18] have performed atom probe tomography (APT) on $Al_{88}Y_7Fe_5$ and found gradients of Al, Y and Fe in concentration profiles, which they believe is an indication of phase separation. Other APT measurements on binary Al-Sm do not show large scale composition fluctuations [19, 20]. Additionally, the final microstructures do not resemble those normally observed during solid state spinodal reactions.

Regardless of mechanism, the genesis of primary crystallization must depend in some way on the structure of the Al-based metallic glass. A dense random packing (DRP) [21] was used to explain the base glass atomic structure of metal-metalloid glasses, but it is not consistent with experiment for Al-based glasses. Anomalous X-ray diffraction of $Al_{87}Y_8Ni_5$ [22], and neutron diffraction of $Al_{87}Ni_7Nd_6$[23] indicate bond shortening between Al and solute atoms, which implies a strong interaction between the solvent atoms and solute atoms. Miracle and Senkov [24-26] proposed the solute-centered cluster concept, and the efficient cluster packing (ECP) of these clusters extends this model to longer length scales. In their model, in order to achieve ECP, there are several types of polyhedral clusters in the atomic structure. Sheng *et al*.[27] used computer simulation coordinated with experiments to investigate the atomic structure and found three types of cluster packing, which are face-sharing, edge-sharing, and vertex sharing clusters. Although these results give insight into the as-quenched atomic structure in Al-based glasses, none of them shows an obvious connection with primary crystallization in Al-based glass. Clearly, a coordinated study of the local atomic structure and the relationship of this structure to the nanocrystallization kinetics is necessary to advance the understanding of primary crystallization in amorphous Al alloys.

In the current study fluctuation electron microscopy (FEM) is used to measure the structure that drives primary crystallization in as-quenched state of Al-based glasses. The FEM technique is sensitive to many-body atom correlations. Therefore, it can be used to detect nanometer-scale structure called medium-range-order (MRO) in amorphous materials[28]. FEM samples spatial fluctuation in diffraction at nm resolution. The fluctuations are quantified by using the normalized variance,

$$V(k,R) = \frac{\langle I^2(k,R,\mathbf{r})\rangle}{\langle I(k,R,\mathbf{r})\rangle^2} - 1 \qquad (1)$$

where $I$ is the annular average of DP in each image, $k$ is the magnitude of scattering vector, $R$ is the resolution, $\mathbf{r}$ is the position over the sample, and the angular bracket means average over many positions. In the post-processing, the annular average diffraction intensity was calculated for each diffraction pattern. Stratton *et al*. [29] first applied this method to Al-based glass, and found FCC Al-like MRO in $Al_{92}Sm_8$. "Al-like" means that this structure diffracts at the Bragg conditions for FCC Al; thus, its internal atomic structure is similar to FCC Al. As shown below and elsewhere [30], the structure may be strained, distorted, or defective. Simulation [29] shows that nanoscale Al order reproduces the peak position and relative peak height compared to the experiment, but that various icosahedral structures do not. Similar MRO was not found in a $Al_{92}Sm_8$ sample amorphized by cold rolling. In differential scanning calorimeter (DSC) measurement, primary crystallization was only observed in melt-quenched samples and not in cold-rolled samples [31]. Additionally, an obvious glass transition, $T_g$ and an exothermic signal corresponding to eutectic phase formation were observed in DSC measurement of cold-rolled sample. Thus, the presence of Al-like MRO is correlated with primary crystallization, and these regions were identified as the quenched-in nuclei. The FEM data do not, however, exclude ECP or any other model for the regions between the Al-like MRO regions. Moreover, the simulations in [29] used Al-like MRO with diameter $d$ of 3 nm only, so that they did not extract the size and volume fraction of Al-like MRO. The diameter and volume fraction data are crucial for quantitative nucleation modeling, and for distinguishing quenched-in nuclei from grain coarsening on a structural basis.

Atom probe tomography performed on $Al_{90}Sm_{10}$ [32] supports the existence of nanosize pure Al region in as-quenched sample, although there is no crystallography information in the APT data. FEM experiments by Wen *et al*. [33] and Daulton *et al*. [34] have confirmed MRO in Al-based glass. In [33], it is proposed that the solute-centered quasi-equivalent clusters are the structural building blocks in $Al_{85}Ni_5Y_{10}$ and $Al_{85}Ni_5Y_8Co_2$, and that the MRO structure consists of organized packing of such clusters. The authors do not exclude the possibility of quenched-in nuclei, but neither do they connect any aspect of the structure to primary crystallization. The two Al-based glasses in [33] exhibit a $T_g$ in DSC. In [34], they adopted phenomenological model developed by Stratton and Voyles [35], and used this model to estimate the correlation length of local order.

Stratton and Voyles [35, 36] developed an analytical model of the variance $V$ as a function of MRO size $d$ and volume fraction $\Phi$. In their model, the sample is divided into many columns of equal size $R \times R \times t$, in which $R$ is the resolution and $t$ is the sample thickness. The column is further divided into bins of cubic shape with size $R$, which are randomly occupied by randomly-oriented Al-like MRO clusters. Perfect Al nanocrystals are used to represent Al-like MRO in the model, although the real Al-MRO may have some internal strain or disorder. How many bins will be occupied by the Al-like MRO is determined by the volume fraction of MRO. It is a simple model, but it incorporates the fundamental physical parameters of atomic structure that dominate the variance from Al-based glasses. They found the variance goes up monotonically with $d$, and goes through maximum as a function of $\Phi$. The variance is inversely proportional to $R^2$ and $t$. Within the limits of thin sample required by kinematic scattering, the $R$ and $t$ dependences have been experimentally confirmed [37, 38].

In this model, the only factor that varies with $k$ (*i.e.* from Bragg reflection to Bragg reflection) is the Bragg active fraction $A_{hkl}$. $A_{hkl}$ is the fraction of randomly-oriented population of nanocrystals that will exhibit strong diffraction into one of a family of reflections $\{hkl\}$. $A_{hkl} \approx (1/4)M_{hkl}\Delta\theta$, and $\Delta\theta = (d_{hkl}/d) + 2(\alpha + \beta)$, where $d_{hkl}$ is the plane spacing for the reflection $\{hkl\}$, $d$ is the nanocrystal diameter, $\alpha$ is the pixel size in the diffraction pattern and $\beta$ is the illumination convergence angle [36]. $M_{hkl}$ is the multiplicity of the $\{hkl\}$ family, defined as the number of unique planes $(hkl)$ that belong to the family $\{hkl\}$. For the fcc structure, $M_{111} = 12$, $M_{200} = 6$, and $M_{220} = 12$ again. This means that reflections with different $M_{hkl}$ have a different dependence on $d$ and $\Phi$, so that, in principle, if $V$ for two reflections $hkl$ with different $M_{hkl}$ is known, $d$ and $\Phi$ can be determined.

In the current work the physical insight from the Stratton/Voyles model is adopted with the main focus on analysis of $V_{111}$ and $V_{200}$ in FEM data from Al-based glass. However, the simplifying assumptions made to render the model analytical tractable are too severe. Some improvements to the model have been developed [39], but it is still not sufficient for a quantitative match to the experimental data. Instead, state-of-the-art multislice dynamical diffraction simulations [40] from atomistic models of Al-based glass are employed which allow to the modeling of the full complexity of the real interaction between the electron beam and sample to reliably extract $d$ and $\Phi$ from FEM data.

In the following, the FEM experiments and simulations are described and combined to extract $d$ and $\Phi$ for $Al_{88}Y_7Fe_5$ metallic glass, and glasses with 1 at. % Cu substituted for Y and substituted for Al. Finally, the results are compared with measurements of the volume fraction and nucleation rate of nanocrystals after crystallization., All of the results are demonstrated to be qualitatively consistent with a primary crystallization driven by retained MRO.. They are inconsistent with grain coarsening, and no evidence for phase separation is observed in any of the microscopy. In part II a quantitative nucleation model is developed that connects the retained MRO measured by FEM with the microstructure after primary nanocrystallization.

## 2. Methods

*2.1 Experiment*

Samples of amorphous $Al_{88}Y_7Fe_5$, $Al_{88}Y_6Fe_5Cu_1$, and $Al_{87}Y_7Fe_5Cu_1$ were prepared by rapid quenching in a single wheel melt spinner at a tangential wheel speed of 55 m/s. Each segment of the ribbon used for further experiments was checked by XRD to confirm that it was completely amorphous. Samples were thinned for TEM by electropolishing in 75% methanol +25% nitric acid at around -42 ºC. After electropolishing, the sample was cleaned using trichloroethylene, acetone, and methanol in sequence to remove organic contamination. Subsequently, the sample was subjected to plasma cleaning for 3 minutes and 15 seconds before loading the sample into the STEM. There is no crystallization induced by plasma cleaning for this time period, which was confirmed by taking electron diffraction patterns (DPs) of plasma cleaned sample and only observing broad, fuzzy rings in the DPs.

STEM FEM experiments were performed using FEI Titan at 200 kV, which yielded a substantial improvement over previous TEM FEM experiments [36-38]. The STEM mode is employed to scan t the beam across the sample area of interest. At each position, a nanodiffraction pattern is collected. In order to remove inelastically scattered electrons, energy filtering with 10eV slit width was used in the FEM experiment. The nanodiffraction patterns (DP) were collected on a 2048×2048 US1000 CCD in a Gatan 865 imaging filter. $I(k,R,r)$ is the annular average of each diffraction pattern. The electron transmittance ratio was used to estimate the sample thickness in units of elastic mean free path, which for $Al_{87}Y_7Fe_5Cu_1$ is 84±1 nm [41]. Samples with 0.70 transmittance, or a physical thickness of 29 nm were used. In order to optimize the tradeoff between the probe coherence and acquisition time, a spot size 8 with a 10 μm condenser aperture was selected for the experiments. Spot size 8 corresponds to the source size of 0.46 nm with probe current of 3 pA. Based on previous results[29], a resolution of 1.88 nm was used in the measurements, which at 200 kV is a convergence half angle of 0.81 mrad. During STEM FEM experiments, at least ten thin separate areas were examined in each sample. In each area, 10 by 10 DPs, or 100 DPs were acquired.. The $V(k,R)$ that arises from shot noise was calculated based on the analysis by Fan *et al*.[42] and subtracted from eqn.(1). Thickness filtering was applied as well using the method described by Hwang and Voyles [37].

Both standard bright field (BF) and dark field (DF) TEM imaging were used to characterize the population of nanocrystals after primary crystallization. The BF or DF images were acquired using Philips CM200 Ultra-twin TEM under 200 kV. The images were obtained from at least four areas in each sample to obtain statistically significant counting. The electron beam transmittance was used to estimate the sample thickness. (Because of large objective aperture in Philips CM200, a calibration was established from Titan STEM analysis on the same sample).

*2.2 Simulation*

The simulated $V(k)$ values were obtained from models consisting of Al nanocrystals embedded in a Al DRP matrix as a function of the nanocrystal diameter, volume fraction, and strain state. Comprehensive simulation results are presented elsewhere [30]; the simulations that best match the experiment are discussed here and were used to extract $d + \Phi$ from the experimental data. The models were constructed and used to calculate the variance as follows. Perfect Al nanocrystals were employed to approximate the MRO in the model. The Lennard-Jones potential presented by Zhang and Xia [43] was adopted to build the DRP matrix in the model. Since the density of amorphous Al glass is very close to its crystalline counterpart, the DRP structure was constructed with same density as the Al FCC crystal, $0.0602/Å^3$. A Metropolis Monte Carlo (MC) method was used to relax the DRP structure until the energy is converged, which is established when the system energy change in 50000 random movements is smaller than 0.1%.

After the DRP structure is built, spherical nanocrystals are constructed of the required size with random orientation. The nanocrystals are inserted into the DRP matrix at random positions with random orientation. The atoms in the matrix overlapping with nanocrystals are removed and a few extra atoms are either added or subtracted randomly in the matrix to maintain constant atom number density. Then the matrix atoms were relaxed using the MC method until the system energy is converged without moving the atoms inside the crystal. (Because the crystals are very small, the atoms are in a high energy state, and the crystal structure will be destroyed during relaxation if those atoms are allowed to rearrange.)

In some models, both hydrostatic strain and disorder in the nanocrystals were considered. The strain was applied to the nanocrystal before it was inserted in the matrix. After the atoms in the matrix are relaxed, the atoms in the nanocrystal are relaxed at 300 K while fixing the matrix atoms. A range of $d$ from 1.20 nm to 1.65 nm corresponding to 55 to 135 atoms inside the nanocrystals was simulated as well as a range of $\Phi$ from 0.0335 to 0.2343 for perfect crystals. A range of $d$ from 1.62 nm to 2.44 nm, and $\Phi$ from 0.0669 to 0.2343 was also simulated for crystals including strain and disorder. To investigate the model size effect on the variance, model sizes varying from 8.44 nm to 18.56 nm at constant $d$ and $\Phi$ were simulated. (The model is a cubic, and the model size refers to cubic edge length).

The variance of this atomic structure was calculated by computing many $I(k, \mathbf{r})$ in different orientations of the model, assuming global structural isotropy. A state-the-art multislice dynamical diffraction algorithm [44] was used to calculate $I(k, \mathbf{r})$. There is a background in $V(k)$ that arises from DRP matrix, which was subtracted.

Each model contains a limited number of nanocrystals, typically 300. In order to check the effect of finite sampling of the random orientations and random positions of the embedded nanocrystals, we generated five different instances of the models with $d$=1.36 nm but different $\Phi$. Then, the variance of the created atomic structure was calculated as a function of $\Phi$. For all the structures, the standard deviation of the variance is within 5% of the mean, which provides an estimate of the statistical uncertainty in the simulated variance of ±5%.

# 3. Results

## 3.1 Experiments

Fig. 1 compares the variance of $Al_{88}Y_7Fe_5$, $Al_{88}Y_6Fe_5Cu_1$ and $Al_{87}Y_7Fe_5Cu_1$ as-spun samples. We have subtracted the matrix background using a Lorentzian function. All the traces exhibit a major peak at 0.39 Å$^{-1}$ with a shoulder on the high-$k$ side near 0.47 Å$^{-1}$. The height of the main peak, the height of the shoulder, and the ratio of the two heights changes with composition. Since the sample thickness in the three alloy samples was controlled carefully to 29.0±2.7 nm, there is a significant difference among the magnitude of the major peak in the variance. In addition, the shoulder is different among these three alloys.

The main peak is assigned to Al {111} diffraction and the shoulder to Al {200}. To decompose these contributions, the first peak region was fitted to a sum of two Gaussians, one for the peak and the other one for the shoulder. The two Gaussian functions were required to have the same width. The FEM measurement and fitting results are shown in Fig. 2. The amplitude and its uncertainty for each Gaussian function are given in Table 1. Fig. 3 shows the volume fraction of nanocrystal change as a function of annealing time. The volume fraction of nanocrystal increases as annealing time increases for all compositions.

## 3.2 Simulation

Fig. 4 shows the simulated variance of the DRP matrix with no nanocrystals and a model of $Al_{89}La_6Ni_6$ constructed by Sheng et al. [27] using molecular dynamics quenching with an empirical potential. The $Al_{89}La_6Ni_6$ model consists of solute-centered quasi-equivalent clusters. The Voronoi polyhedra that define those clusters are not connected in an ordered fashion. Fig. 4 shows that $V(k)$ from the DRP matrix and the MD model are virtually the same. The small oscillations on top of the sloping

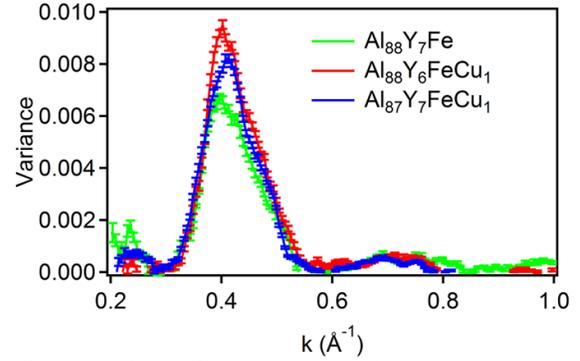

Fig. 1: Variance for as-spun samples.

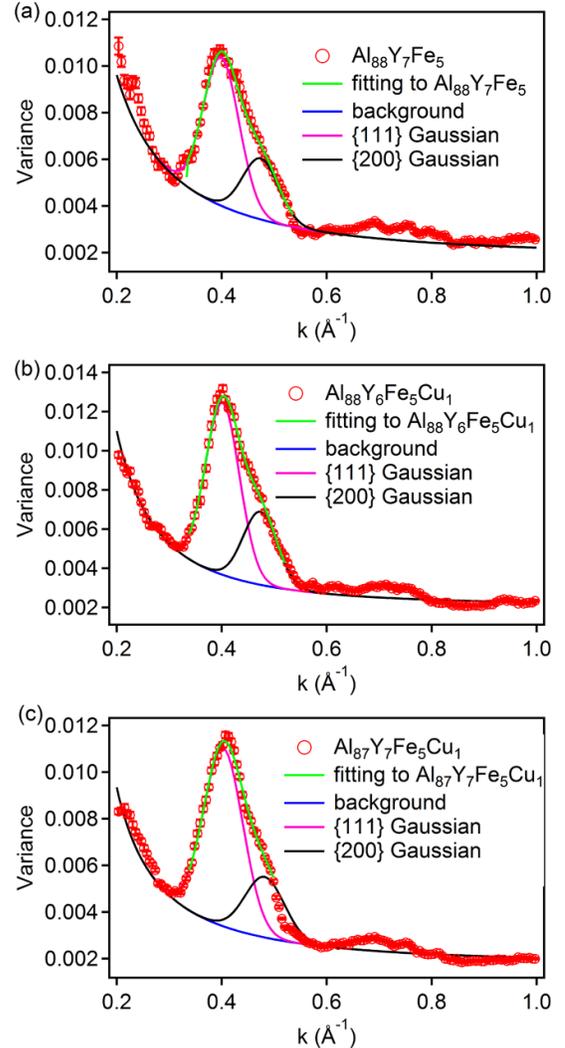

Fig. 2 (a) Fitting results for $Al_{88}Y_7Fe_5$ (b) $Al_{88}Y_6Fe_5Cu_1$, and (c) $Al_{87}Y_7Fe_5Cu_1$. The green line are the fitting to the major peak, the blue line is the background using Lorentzian function, the pink line is the Gaussian peak due to {111} Bragg reflection, and the black line is the Gaussian peak due to {200} Bragg reflection.

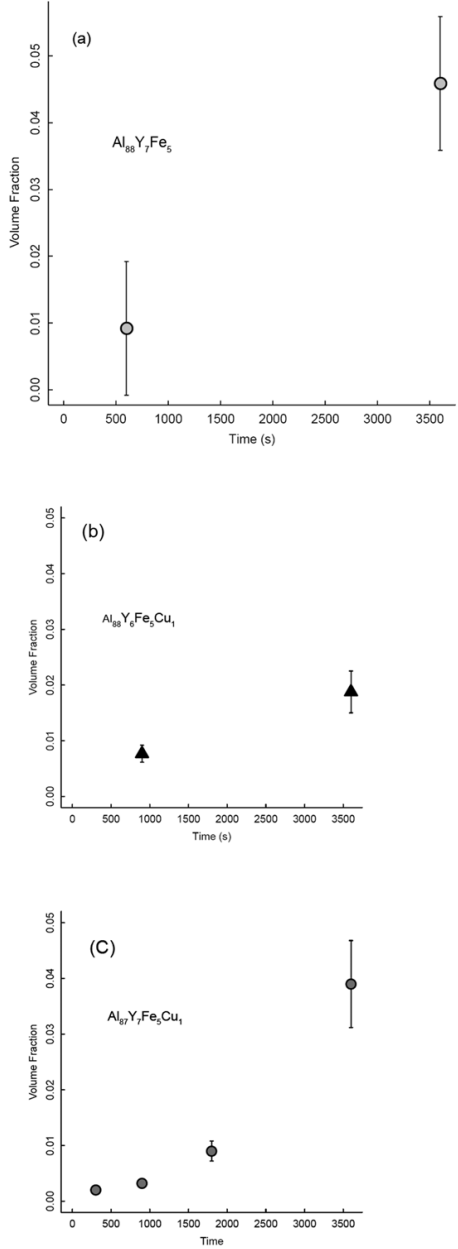

Table 1 Fitting parameters for $Al_{88}Y_7Fe_5$, $Al_{88}Y_6Fe_5Cu_1$ and $Al_{87}Y_7Fe_5Cu_1$.

|  | {111} Gaussian peak | {200} Gaussian peak |
|---|---|---|
| $Al_{88}Y_7Fe_5$ | 0.0064±0.001 | 0.0026±0.0002 |
| $Al_{88}Y_6Fe_5Cu_1$ | 0.0088±0.0002 | 0.0037±0.0002 |
| $Al_{87}Y_7Fe_5Cu_1$ | 0.0076±0.0004 | 0.0026±0.0003 |

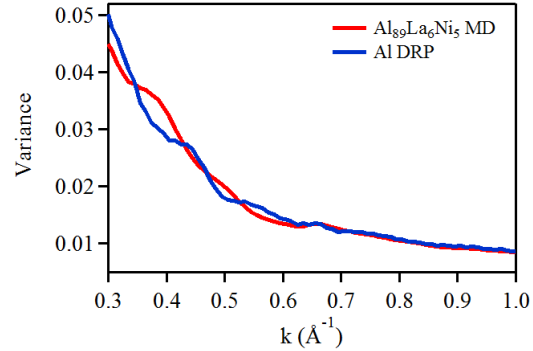

Figure 4: Simulated variance of the DRP structure used as the matrix for the simulations presented here, and a model $Al_{89}La_6Ni_5$ consisting of quasi-equivalent clusters, created by molecular dynamics quenching [27].

Figure 3 Volume fraction of nanocrystal change as a function of annealing time (a) $Al_{88}Y_7Fe_5$ (b) $Al_{88}Y_6Fe_5Cu_1$ and (c) $Al_{87}Y_7Fe_5Cu_1$. Both $Al_{88}Y_7Fe_5$ and $Al_{87}Y_7Fe_5Cu_1$ were annealed at 245 °C, and $Al_{88}Y_6Fe_5Cu_1$ was annealed at 200 °C.

background found in both $V(k)$ curves is likely to an artifact of the limited model size.

The variance as a function of $k$ for representative models containing different sizes and volume fractions of nanocrystals are shown Fig. 5. In these models, the embedded nanocrystals are perfect crystals. Fig. 5 shows that variance for {111} and {200} both generally increases as a function of $d$ and $\Phi$.

The experimental peak positions are shifted systematically to lower $k$ compared to the perfect Al Bragg reflections, suggesting that the embedded Al-like regions are under tensile strain. The hydrostatic tensile strain systematically shifts the peaks in $V(k)$ [30], and the strain-induced disorder suppresses peaks at high $k$ [30].

From simulations for a series of strains, a 4.1% tensile hydrostatic strain in the nanocrystal was found to give the best match to experiment for both the magnitude of the variance and the peak position. Fig. 6 shows the variance for the model with 4.1% tensile strain and disorder by relaxing atoms in the nanocrystal, without allowing the nanocrystal volume to change. $V_{111}$ and $V_{200}$ generally increase as a function of $d$ and $\Phi$, but the magnitude of peaks is suppressed by disorder. This level of disorder effectively suppresses the

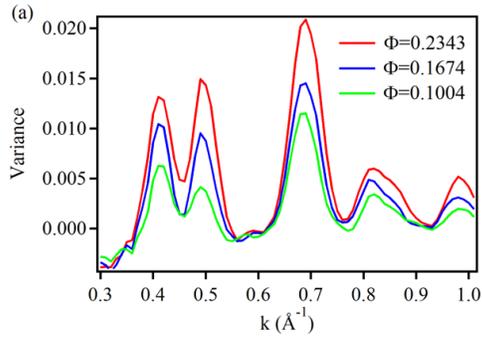
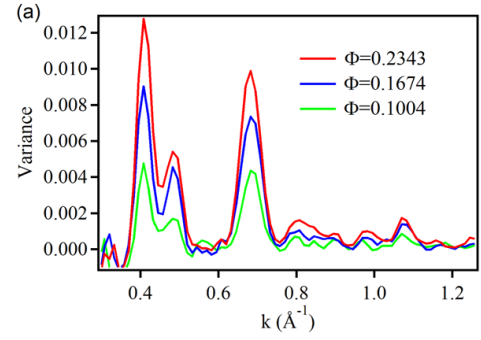
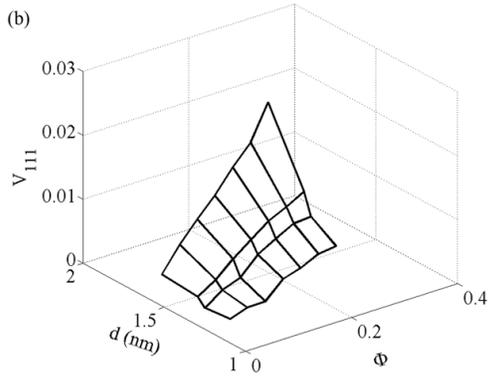
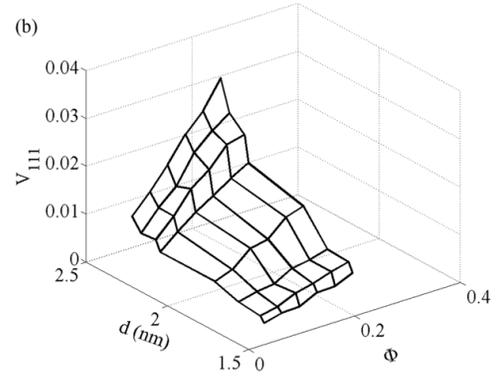
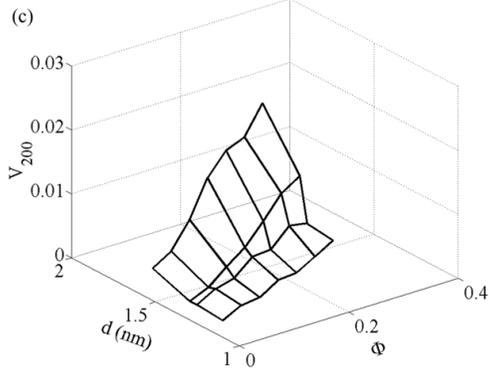
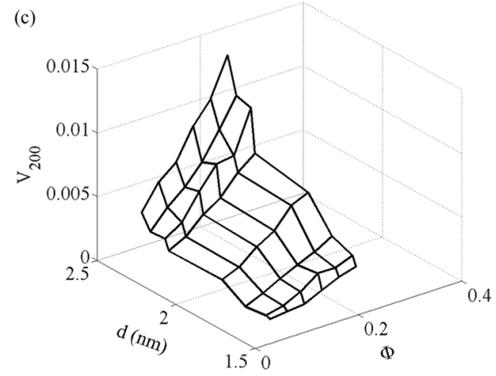

Figure 5 For number density ρ=0.06/Å$^3$, $t$=118.12 Å and $R$=18.8 Å (a) Variance for $d$ = 14 Å at different volume fraction Φ, and Variance as a function of nanocrystal size $d$ and volume fraction Φ for (b) {111} plane and (c) {200} plane using femauto algorithm.

Figure 6 For number density ρ=0.06/Å$^3$, and $R$=18.8 Å (a) Variance for $t$=168.74 Å and $d$=16.2 Å at different volume fraction Φ, and Variance as a function of nanocrystal size $d$ and volume fraction Φ for (b) {111} plane and (c) {200} plane using femauto algorithm. Both (b) and (c) are normalized to $t$=29 nm. There is 4.1% hydrostatic strain in the nanocrystal, and the atoms in the nanocrystals are relaxed at 300 K.

Al {220} peak at 0.67 Å$^{-1}$, as seen experimentally in Fig. 1. The local maxima and minima are noise in simulations due to the limited number of nanocrystals in the model. In the analytical simulation, the variance is third order in $r$. Therefore, at constant Φ, a polynomial function of form $f(r) = a_1 \times r^3 + a_2 \times r^2 + a_3 \times r$ was used to fit $V$ as a function of $r$ at constant Φ based on non-linear least square method. This functional form is based on the Stratton and Voyles analytical model results [36]. The fit (and therefore smoothed) $V_{111}(r, \Phi)$ and $V_{200}(r, \Phi)$ are shown in Fig. 7, where a linear interpolation was employed to fill the space at finer grid ($\Phi$).

## 4. Discussion

### 4.1 The implication from simulation results

The agreement of the simulated $V(k)$ between the DRP matrix of the current models and the MD model of $Al_{89}La_6Ni_5$ in Fig. 4 justifies our neglect of the Y and Fe atoms in the matrix of our model, since including atoms of similar scattering power does not change the simulated $V$. Neither model agrees with the experimental data in Fig. 1 or Fig. 2, so the MD model does not capture the structure of the real material. As shown by the simulations in Fig. 5, fcc-Al-like nanoscale order is the missing ingredient in the MD model. These results also show that FEM signal is dominated by the nanoscale Al-like regions, which, in this system, renders it insensitive to the structural difference between the pure Al DRP and the MD model. The FEM data are therefore silent on the question of

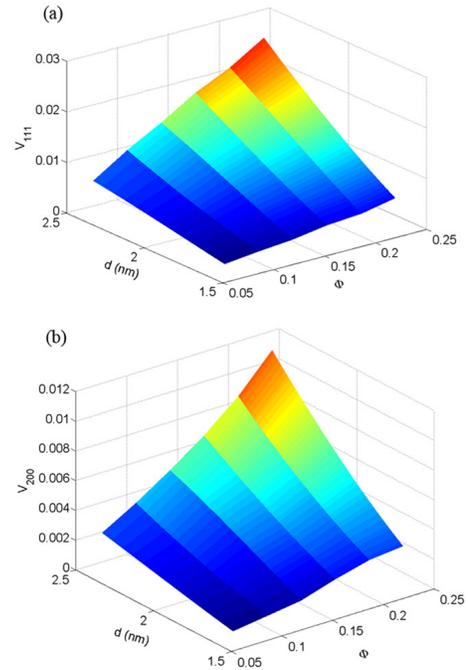

Figure 7 at $\varepsilon=4.1\%$ (a) $V_{111}$ as a function of $d$ and $\Phi$ (b) $V_{200}$ as a function of $d$ and $\Phi$.

what structure lies between the Al-like regions. It may be quasi-equivalent clusters as in the MD model, or it may not be. This insensitivity to subtle details in quasi-equivalent clustering but high sensitivity to crystal-like order has been noted previously by Wen *et al.* [45].

The variance of Bragg reflections intensity at both {111} and {200} positions increases monotonically as a function of nanocrystal size $d$ and volume fraction $\Phi$ in the calculation range. In the Stratton/Voyles model, $V(\Phi)$ reaches a maximum at fairly low $\Phi$, less than 10%, then decreases as $\Phi$ continues to increase [35, 36]. This maximum is not observed in the simulations, up to $\Phi$ of 20%. Qualitatively, this behavior must be correct: A perfect single crystal, for which $\Phi = 1$ and $A_{hkl} = 1$ must have $V = 0$, since the structure and thus the diffracted intensity are the same everywhere. However, the Stratton/Voyles model severely underestimates the $\Phi$ at which this occurs. This is the result of the simplifying assumptions that omit several sources of variability in the diffracted intensity from an ensemble of nanoparticles which are captured in the simulations in Fig. 5 and 6. This is explored in greater detail elsewhere [30], but the most important factor that was not included is the detailed dependence of the diffracted intensity on the deviation parameter, $s$. that is the distance of the reciprocal lattice point corresponding to a reflection ($hkl$) from the Ewald sphere. In the kinematic limit, the diffracted intensity oscillates as a function of $s$ [46]. In the Voyles/Stratton model, the diffraction from a given nanocrystal is either on or off. The simulations also capture disregistry between the positions of the probes and the positions of the nanocrystals, which is not allowed in the Stratton/Voyles model. If the probe sometimes catches just a small portion of a nanocrystal, but other times illuminates the entire nanocrystal, then the variability in diffraction intensity will increase from place to place and act to increase $V$. Finally, the Stratton/Voyles model predicts $V = 0$ for the DRP matrix, which is not observed in the simulations. Thus, the simple assumption that $I \propto N$, where $N$ is the number of matrix atoms under the probe must not be correct.

The remaining discrepancy between the simulations and experiment is that the width of the peaks in $V(k)$ in the experiment is wider than in simulation. The peak width depends on instrumental factors such as the probe convergence angle and on the state of structural disorder in the sample, with more disorder leading to broader peaks. Disorder also changes the peak magnitude, however, and if a sufficiently large disorder is introduced to match the experimental peak widths, $V_{200}$ drops to nearly zero, which does not agree with the experiments. Since only the peak magnitudes are employed for the current analysis, further refinements were not used to match the peak width exactly.

The disorder associated with strain decreases $V$ at all $k$, but the decrease is larger at higher $k$. This is because disorder suppresses diffraction more for small plane spacings and high $k$. Similar results were found in models of amorphous silicon with embedded crystal-like regions [47]. A perfect Si crystal results in much high variance at high $k$ value, and using strained paracrystalline Si removed this effect.

*4.2 Extracting (d, Φ) from experiments*

Graphically, $(d, \Phi)$ can be extracted from the experimental data by the following steps. First, the experimental values for $V_{111}$ and $V_{200}$ define $(d, \Phi)$ contours on the surface plots in Fig. 7. The intersection of these two contours gives a single $(d, \Phi)$ for a given $(V_{111}, V_{200})$. The best estimate of $d$, $\Phi$ and the uncertainty of $r$ and $\Phi$ arising from uncertainty in $V_{111}$ and $V_{200}$ are calculated using a Monte Carlo approach. First, $1.6 \times 10^5$ $(V_{111}^G, V_{200}^G)$ pairs were created with a Gaussian probability density function with the mean and standard deviation given by the experimental values in the Table 1. Then, $V_{111}(d, \Phi)$ and $V_{200}(d, \Phi)$ are searched to find $(d, \Phi)$ numerically in Fig. 7, which generates $1.6 \times 10^5$ $(d, \Phi)$ pairs. 2D histograms of the $(d, \Phi)$ lists are shown in Fig. 8. Lastly, the 2D Gaussian function of the form

$$A_0 + A_1 \exp\left[\frac{-1}{2(1-cor^2)}\left[\left(\frac{d-\bar{d}}{\sigma_d}\right)^2 + \left(\frac{\Phi-\bar{\Phi}}{\sigma_\Phi}\right)^2 - \frac{2cor\left(d-\bar{d}\right)\left(\Phi-\bar{\Phi}\right)}{\sigma_r\sigma_\Phi}\right]\right] \quad (2)$$

was used to fit each histogram in Fig. 8. Table 2 lists $\bar{d}$ and $\bar{\Phi}$ as the best estimates for each composition, and $\sigma_d$ and $\sigma_\Phi$ as the uncertainties. However, this ignores the correlation between the uncertainties in $d$ and $\Phi$, which is shown graphically by the contour on each figure, which are drawn at a value of $1/e$ of the maximum. The shape of the histograms shows that at large confidence intervals, the uncertainties of $r$ and $\Phi$ are strongly correlated. However, the maximum of each histogram is well fitted by the 2D Gaussian model.

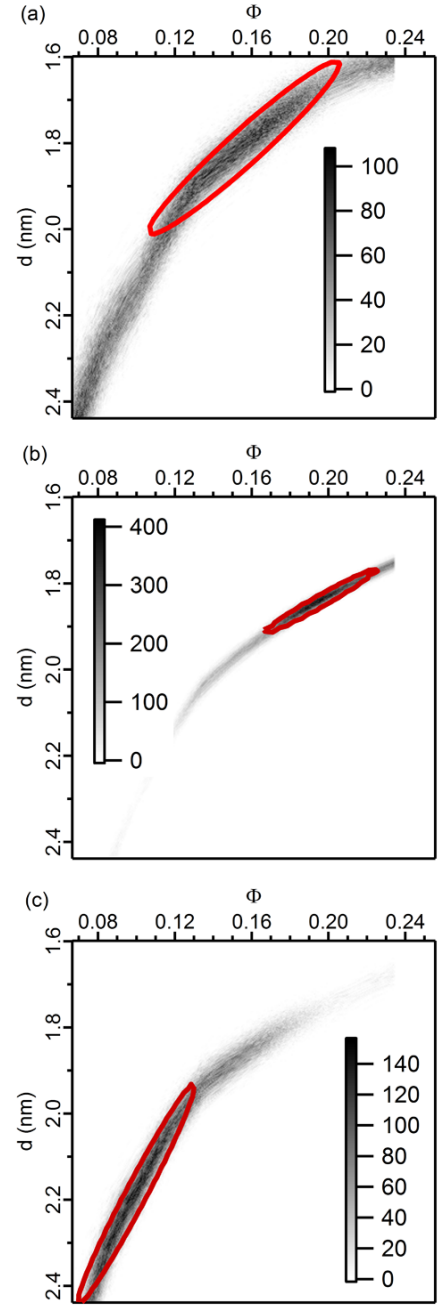

Fig. 8 The histogram for (a) $Al_{88}Y_7Fe_5$ (b) $Al_{88}Y_6Fe_5Cu_1$ and (c) $Al_{87}Y_7Fe_5Cu_1$. The red line in the figure is contour of fitted 2D Gaussian function with value $A_1/e$.

*4.3 Connection with $T_x$ and post annealing measurement*

The experimental primary crystallization temperature $T_x$ from DSC [14] and the nanocrystal density after annealing treatment from TEM for these three alloys are shown in Table 3. Both $Al_{88}Y_7Fe_5$ and $Al_{88}Y_6Fe_5Cu_1$ were annealed at 245 ºC for 1 hour, and $Al_{87}Y_7Fe_5Cu_1$ was annealed at 200 ºC for 1 hour. $T_x$ represents the onset of detectable nucleation and is determined by three factors: the density of nucleation sites, the potency of the nucleation sites, and the transient nucleation delay time, which is governed by the relevant diffusivities. As shown in Table 3, the diameter of the nuclei are almost constant from alloy to alloy, within the experimental uncertainty. The density of MRO calculated from Table 2 based on $\rho_{MRO} = \Phi_{MRO}/(\pi d^3/6)$ is also shown for comparison. As indicated in Table 3, the rank order of $\rho_{MRO}$ is the same as that of $\rho_{crystal}$, and opposite to the rank order of $T_x$, although the $\rho_{MRO}$ are not really different outside of experimental uncertainty between $Al_{88}Y_7Fe_5$ and $Al_{88}Y_6Fe_5Cu_1$. This is due entirely to the uncertainty in $d$ for the base alloy. The rank order of the volume fraction $\Phi$ is distinguishable and follows $\rho_{crystal}$ directly. This correlation indicates that MRO acts as heterogeneous site for nanocrystal nucleation. During continuous heating, if there are more MRO clusters and thus nucleation sites in the as-quenched sample, the nucleation rate is higher, shifting $T_x$ lower. During isothermal annealing, higher MRO density results in higher crystal density by providing a higher density of nucleation sites. However, the difference in $\Phi$ between the base alloy ($Al_{88}Y_7Fe_5$) and 1 at.% substituted for Al ($Al_{87}Y_7Fe_5Cu_1$) is much smaller than the change in $T_x$.

In addition, Fig. 3 demonstrates that the nanocrystal volume fraction in the base alloy and the alloys with Cu substitution increases as the annealing time increases. If the grain coarsening mechanism were valid for growth, the volume fraction of nanocrystal should be constant during annealing treatment. However, Fig. 3 indicates this is not correct. Moreover, Φ in Table 2 is much higher than the crystal volume fraction at the shortest time in Fig. 3. If Φ in the as-quenched state is treated as the initial crystal fraction in an amorphous/nanocrystal composite, then a larger drop in Φ should occur at the start of the reaction. This is also inconsistent with grain coarsening.

Table 2 $d$ and $\Phi$ of MRO in $Al_{88}Y_7Fe_5$, $Al_{88}Y_6Fe_5Cu_1$ and $Al_{87}Y_7Fe_5Cu_1$.as-spun sample.

|  | $d$(nm) | $\Phi$ |
|---|---|---|
| $Al_{88}Y_7Fe_5$ | 1.8±0.1 | 0.16±0.03 |
| $Al_{88}Y_6Fe_5Cu_1$ | 1.84±0.06 | 0.20±0.02 |
| $Al_{87}Y_7Fe_5Cu_1$ | 2.2±0.2 | 0.10±0.02 |

Table 3 $T_x$, $\rho_{crystal}$ and $\rho_{MRO}$ for $Al_{88}Y_7Fe_5$, $Al_{88}Y_6Fe_5Cu_1$ and $Al_{87}Y_7Fe_5Cu_1$.as-spun sample.

|  | $\rho_{MRO}(10^{25}/m^3)$ | $\rho_{nanocrystal}(10^{21}/m^3)$ | $T_x$ (ºC) |
|---|---|---|---|
| $Al_{88}Y_7Fe_5$ | 5.2±1.0 | 3.5±1 | 265±5 |
| $Al_{88}Y_6Fe_5Cu_1$ | 6.1±0.6 | 13±4 | 215±5 |
| $Al_{87}Y_7Fe_5Cu_1$ | 1.8±0.4 | 0.7±0.2 | 275±5 |

There is no evidence of phase separation in any of the data. The Z-contrast STEM images that were used as a guide for the STEM FEM experiments [37] are sensitive to composition and thickness changes. In several hundred images, a thickness gradient is detected in towards the holes in the samples introduced by electropolishing, but not the patterned structures proposed for phase separation based on atom probe tomography. Moreover, the circular arrangements of nanoparticles seen by Gangopadhyay [17] were not observed in any of the conventional TEM images.

## 5. Conclusion

By combining FEM experiments and simulations, the diameter and volume fraction of retained MRO regions were determined in $Al_{88}Y_7Fe_5$, $Al_{87}Y_7Fe_5Cu_1$ and $Al_{88}Y_6Fe_5Cu_1$ metallic glasses. Comparing the diameter and volume fraction in the as-quenched state to the crystallization temperature and nanocrystal volume fraction after devitrification for all three alloys strongly supports the activity of MRO regions to catalyze the primary crystallization. The data are inconsistent with a grain coarsening model and reveal no evidence for a precursor phase separation reaction.

## Acknowledgement

Work by F.Y. and P.M.V. was supported by the U.S. National Science Foundation (DMR-0905793). Work by S.D.I. and J.H.P. was support by . NSF (DMR-1005334). The authors thank H. W. Sheng for sharing the coordinates of his $Al_{89}La_6Ni_5$ structural model.

## References

[1] Y. He, S.J. Poon, G.J. Shiflet, Synthesis and properties of metallic glasses that contain aluminum, Science 241 (1988) 1640-1642.
[2] G.J. Shiflet, Y. He, S.J. Poon, Mechanical properties of a new class of metallic glasses based on aluminum, J. Appl. Phys. 64(12) (1988) 6863-6865.
[3] J.C. Foley, D.R. Allen, J.H. Perepezko, Strategies for the development of nanocrystalline materials through devitrification, Mater. Sci. & Eng. A 226-228 (1997) 569-573.
[4] J.C. Foley, D.R. Allen, J.H. Perepezko, Analysis of nanocrystal development in Al-Y-Fe and Al-Sm glasses, Scripta. Mater. 35(5) (1996) 655-660.
[5] Y.-H. Kim, A. Inoue, T. Masumoto, Increase in mechanical strength of Al-Y-Ni amorphous alloys by dispersioin of nanoscale fcc-Al particles, MATER Trans. JIM 32(4) (1991) 331-338.
[6] Z.C. Zhong, X.Y. Jiang, A.L. Greer, Nanocrystallization in Al-based amorphous alloys, Phil. Mag.B. 76(4) (1997) 505-510.
[7] A. Inoue, Amorphous, nanoquasicrystalline and nanocrystalline alloys in Al-based systems, Prog. Mat. Sci. 43 (1998) 365-520.
[8] J.T. Fan, Z.F. Zhang, F. Jiang, J. Sun, S.X. Mao, Ductile to brittle transition of $Cu_{46}Zr_{47}Al_7$ metallic glass composites, Mater. Sci. & Eng. A 487 (2008) 144-151.

[9] T.M. Heil, K.J. Wahl, A.C. Lewis, J.D. Mattison, M.A. Willard, Nanocrystalline soft magnetic ribbons with high relative strain at fracture Appl. Phys. Lett. 90 (2007) 212508.
[10] Y. He, H. Chen, G.J. Shiflet, S.J. Poon, On the structural nature of aluminum-based metallic glasses, PHil. Mag.Lett. 61(5) (1990) 297-303.
[11] M. Blank-Bewersdorff, Crystallization behaviour of $Al_{86}Ni_{10}Zr_4$ and $Al_{86}Fe_{10}Zr_4$ metallic glasses, J. Mater. Sci. Lett. 10 (1991) 1225.
[12] R.F. Cochrane, P. Schumacher, A.L. Greer, Crystallization of amorphous $Al_{85}Ni_{10}Ce_5$ alloy, Mater. Sci. Eng. A 133 (1991) 367-370.
[13] K. Nakazato, Y. Kawamura, A.P. Tsai, A. Inoue, On the growth of nanocrystalline grains in an aluminum-based amorphous alloy, Appl. Phys. Lett. 63(19) (1993) 2644-2646.
[14] S. Imhoff, Glass stability and nanocrystal formation in aluminum-based systems, University of Wisconsin-Madison, 2010.
[15] J.H. Perepezko, R.J. Hebert, Amorphous aluminum alloys-synthesis and stability, JOM 54 (2002) 34-39.
[16] L.Q. Xing, A. Mukhopadhyay, W.E. Buhro, K.F. Kelton, Improved Al-Y-Fe glass formation by microalloying with Ti, Phil. Mag. Lett. 84(5) (2004) 293-302.
[17] A.K. Gangopadhyay, T.K. Croat, K.F. Kelton, The effect of phase separation on subsequent crystallization in $Al_{88}Gd_6La_2Ni_4$, Act. Mater. 48 (2000) 4035-4043.
[18] K.K. Sahu, N.A. Mauro, L. Longstreth-Spoor, D. Saha, Z. Nussinov, M.K. Miller, K.F. Kelton, Phase separation mediated devitrification of $Al_{88}Y_7Fe_5$ glasses, Act. Mater. 58 (2010) 4199-4206.
[19] Y.E. Kalay, L.S. Chumbley, M.J. Kramter, I.E. Anderson, Local structure in marginal glass forming Al-Sm alloy, Intermetallics 18 (2010) 1676-1682.
[20] Y.E. Kalay, C. Yeager, L.S. Chumbley, M.J. Kramter, I.E. Anderson, Initial crystallization in a nanostructured Al–Sm rare earth alloy, J. Non-Cryst. 356 (2010) 1416-1424.
[21] J.D. Bernal, J. Mason, Packing of Spheres: Co-ordination of Randomly Packed Spheres, Nature 188 (1960) 910-911.
[22] E. Matsubara, Y. Waseda, A. Inoue, H. Ohtera, T. Masumolo, Anomalous X-ray scattering on amorphous Al87Y8Ni5 and Al90Y10 alloys, Z Naturaforsch A 44 (1989) 814-820.
[23] K. Ahn, D. Louca, S.J. Poon, G.L. Shiflet, Local structure of Al- and Fe-based metallic glasses, J.Phys.: Condens. Matter. 15 (2003) 2357-2364.
[24] D.B. Miracle, O.N. Senkov, A geometrical model for atomic configurations in amorphous Al alloys, J. Non-Cryst. Solids 319 (2003) 174-191.
[25] D.B. Miracle, A structural model for metallic glasses, Nature Mat. 3 (2004) 697-702.
[26] D.B. Miracle, The efficient cluster packing model-An atomic structural model for metallic glasses, Acta Mater. 54 (2006) 4317-4336.
[27] H.W. Sheng, Y.Q. Cheng, P.L. Lee, S.D. Shastri, E. Ma, Atomic packing in multicomponent aluminum-based metallic glasses, Acta. Mater. 56 (2008) 6264-6272.
[28] M.M. Treacy, J.M. Gibson, L. Fan, D.J. Paterson, I. Mcnulty, Fluctuation microscopy: a probe of medium range order, Rep. Prog. Phys. 68 (2005) 2899-2944.
[29] W.G. Stratton, J. Hamann, J.H. Perepezko, P.M. Voyles, X. Mao, S.V. Khare, Aluminum nanoscale order in amorphous $Al_{92}Sm_8$ measured by fluctuation electron microscopy, Appl. Phys. Lett. 86 (2005) 141910-1-141910-3.
[30] F. Yi, P.M. Voyles, Analytical and computational modeling of fluctuation electron microscopy from a nanocrystal/amorphous composite, Ultramicroscopy 122 (2012) 37-47.
[31] G. Wilde, H. Sieber, J.H. Perepezko, Glass formation versus nanocrystallization in an $Al_{92}Sm_8$ alloys, Script Mater. 40(7) (1999) 779-783.
[32] Y.E. Kalay, L.S. Chumbley, I.E. Anderson, Crystallization behavior in a highly driven marginal glass forming alloy, J. Non-Cryst. Solids 354 (2008) 3040-3048.
[33] J. Wen, H.W. Yang, H. Guo, B. Wu, M.L. Sui, J.Q. Wang, E. Ma, Fluctuation electron microscopy of Al-based metallic glasses: effects of minor alloying addition and structural relaxation on medium-range structural homogeneity, J. Phys.: Condens. Matter 19(45) (2007) 455211.

[34] T.L. Daulton, K.S. Bondi, K.F. Kelton, Nanobeam diffraction fluctuation electron microscopy technique for structural characterization of disordered materials — Application to $Al_{88-x}Y_7Fe_5Ti_x$ metallic glasses, Ultramicroscopy 110 (2010) 1279-1289.
[35] W.G. Stratton, P.M. Voyles, A phenomenological model of fluctuation electron microscopy for a nanocrystal/amorphous composite, Ultramicroscopy 108 (2008) 727-736.
[36] W.G. Stratton, P.M. Voyles, Comparison of fluctuation electron microscopy theories and experimental methods, J.Phys.: Condens. Matter. 19 (2007) 455203.
[37] J. Hwang, P.M. Voyles, Variable resolution fluctuation electron microscopy on Cu-Zr metallic glass using a wide range of coherent STEM probe size, Microsc Microanal 17 (2011) 67.
[38] F. Yi, P.M. Voyles, Accepted by Ultramicroscopy.
[39] F. Yi, Medium range order in Al-based metallic glasses, University of Wisconsin-Madison, Madison, 2011, p. 118.
[40] E.J. Kirkland, Advanced computing in electron microscopy, Plenum Press, New York, 1998.
[41] D. Schewiss, J. Hwang, Unpublished work.
[42] L. Fan, D.J. Paterson, I. Mcnulty, M.M. Treacy, J.M. Gibson, Fluctuation X-ray microscopy: a novel approach for the structural study of disordered materials, J. Microsc. 225 (2007) 41-48.
[43] H. Zhang, Z.N. Xia, Molecular dynamics simulation of cluster beam Al deposition on Si (1 0 0) substrate, NUCL INSTRUM METH B 160(3) (2000) 372-376.
[44] E.J. Kirkland, Advanced Computing in Electron Microscopy, Plenum Press, New York, 1998.
[45] J. Wen, Y.Q. Cheng, J.Q. Wang, E. Ma, Distinguishing medium-range order in metallic glasses using fluctuation electron microscopy: A theoretical study using atomic models, J. Appl. Phys. 105 (2009) 043519.
[46] E. Rossmanith, Kinematical intensity profiles obtained for single and multiple diffraction in perfect spherical crystals, J APPL CRYSTALLOGR 33 (2000) 323-329.
[47] S.N. Bogle, P.M. Voyles, S.V. Khare, J.R. Abelson, Quantifying nanoscale order in amorphous materials: simulating fluctuation electron microscopy of amorphous silicon, J. Phys.: Condens. Matter 19 (2007) 455204.